\documentclass[fleqn,10pt]{wlscirep}
\usepackage[utf8]{inputenc}
\usepackage[T1]{fontenc}
\usepackage{lineno}
\title{Metropolitan Scale and Longitudinal Dataset of Anonymized Human Mobility Trajectories}

\author[1,3,$\dag$,*]{Takahiro Yabe}
\author[2,$\dag$,*]{Kota Tsubouchi}
\author[2]{Toru Shimizu}
\author[3]{Yoshihide Sekimoto}
\author[3]{Kaoru Sezaki}
\author[1,5]{Esteban Moro}
\author[1,4]{Alex Pentland}
\affil[1]{Institute for Data, Systems, and Society, Massachusetts Institute of Technology, Cambridge, MA 02139, USA}
\affil[2]{Yahoo Japan Corporation, Kioicho, Tokyo 102-8282, Japan}
\affil[3]{Center for Spatial Information Science,the University of Tokyo, Kashiwa, Chiba 277-8568, Japan}
\affil[4]{Media Lab, Massachusetts Institute of Technology, Cambridge, MA 02139, USA}
\affil[5]{Grupo Interdisciplinar de Sistemas Complejos (GISC), Departamento de Matemáticas, Universidad Carlos III de Madrid, 28911 Leganés, Madrid, Spain}

\affil[*]{corresponding authors: Takahiro Yabe (tyabe@mit.edu) and Kota Tsubouchi (ktsubouc@yahoo-corp.jp)}

\affil[$\dag$]{these authors contributed equally to this work}

\begin{abstract}
Modeling and predicting human mobility trajectories in urban areas is an essential task for various applications. The recent availability of large-scale human movement data collected from mobile devices have enabled the development of complex human mobility prediction models. However, human mobility prediction methods are often trained and tested on different datasets, due to the lack of open-source large-scale human mobility datasets amid privacy concerns, posing a challenge towards conducting fair performance comparisons between methods. To this end, we created an open-source, anonymized, metropolitan scale, and longitudinal (90 days) dataset of 100,000 individuals' human mobility trajectories, using mobile phone location data. The location pings are spatially and temporally discretized, and the metropolitan area is undisclosed to protect users' privacy. The 90-day period is composed of 75 days of business-as-usual and 15 days during an emergency. To promote the use of the dataset, we will host a human mobility prediction data challenge (`HuMob Challenge 2023') using the human mobility dataset, which will be held in conjunction with ACM SIGSPATIAL 2023.
\end{abstract}
\begin{document}

\flushbottom
\maketitle

\thispagestyle{empty}

% \noindent Please note: Abbreviations should be introduced at the first mention in the main text – no abbreviations lists or tables should be included. Structure of the main text is provided below.

\section*{Background \& Summary}

% (700 words maximum) An overview of the study design, the assay(s) performed, and the created data, including any background information needed to put this study in the context of previous work and the literature. The section should also briefly outline the broader goals that motivated the creation of this dataset and the potential reuse value. We also encourage authors to include a figure that provides a schematic overview of the study and assay(s) design. The Background \& Summary should not include subheadings. This section and the other main body sections of the manuscript should include citations to the literature as needed. 

Understanding, modeling, and predicting human mobility trajectories in urban areas is an essential task for various domains and applications, including human behavior analysis \cite{gonzalez2008understanding}, transportation and activity analysis \cite{jiang2017activity}, disaster risk management \cite{yabe2022mobile}, epidemic modeling \cite{oliver2020mobile}, and urban planning \cite{ratti2006mobile}. Traditionally, travel surveys and census data have been utilized as the main source of data to understand such macroscopic urban dynamics \cite{sekimoto2011pflow}. The recent availability of large-scale human movement and behavior data collected from (often millions of) mobile devices and social media platforms \cite{blondel2015survey} have enabled the development and testing of complex human mobility models, resulting in a plethora of proposed methods for the prediction of human mobility traces \cite{luca2021survey}.

Despite its academic popularity and societal impact, human mobility modeling and prediction methods are often trained and tested on different proprietary datasets, due to the lack of open-source and large-scale human mobility datasets amid privacy concerns \cite{de2018privacy}.
This makes it difficult to conduct fair performance comparisons across different methods. 
Several efforts have created open-source datasets of human mobility. Real-world trajectory datasets include the GeoLife dataset, T-Drive trajectory dataset, and NYC Taxi and Limousine Commision dataset. The GeoLife dataset \cite{zheng2008geolife} provides trajectory data of 182 users across a period of over three years, containing 17,621 trajectories with a total distance of about 1.2 million kilometers and a total duration of 48,000 hours. The T-Drive trajectory dataset contains trajectories of 10,357 taxis across a one week timeframe \cite{yuan2010t}. The total number of points in this dataset is about 15 million and the total distance of the trajectories reaches 9 million kilometers. Similarly, the New York City Taxi and Limousine Commission (NYC-TLC) provides pick-up and drop-off locations and timestamps data \footnote{\url{https://www.nyc.gov/site/tlc/about/tlc-trip-record-data.page}}. Although T-Drive and NYC-TLC datasets provide massive amounts of trajectory information, they are limited to taxi trips.
There has also been several synthetic datasets produced from open-source data, including the Open PFLOW \cite{kashiyama2017open} and Pseudo-PFLOW datasets \cite{kashiyama2022pseudo}. 
While such datasets are valuable in conducting large-scale experiments on human mobility prediction, the lack of metropolitan-scale, longitudinal, real-world, and open-source datasets of individuals has been one of the key barriers hindering the progress of human mobility model development. 

To this end, we created an open-source and anonymized dataset of human mobility trajectories from mobile phone location data provided by Yahoo Japan Corporation. 
The dataset contains 100,000 individuals' mobility trajectories across a 90 day period collected from an undisclosed, highly populated metropolitan area in Japan. The location pings are discretized into 500 meters x 500 meters grid cells and the timestamps into 30 minute bins. The actual date of the observations are not available either (i.e., timeslot $t$ of day $d$) to protect privacy. 
The 90 day period is composed of 75 days of business-as-usual and 15 days during an emergency with unusual behavior. 

To promote the use of the dataset, we will host a human mobility prediction data challenge (`HuMob Challenge 2023') using the human mobility dataset of 100K individuals’ trajectories across 90 days. The workshop will be held in conjunction with ACM SIGSPATIAL 2023 \footnote{\url{https://sigspatial2023.sigspatial.org/}}. Participants will be tasked to develop and test methods to predict human mobility trajectories using the provided open-source dataset (for details, see Section `Human Mobility Prediction Data Challenge').

\section*{Methods}

\subsection*{Observation of Smartphone GPS records}
GPS location data were collected from smartphones which have installed the Yahoo Japan Application, and were anonymized so that individuals cannot be specified, and personal information such as gender, age and occupation are unknown. 
Each GPS location record contains the user's unique ID, timestamp of the observation, longitude, and latitude, and the data has a sample rate of approximately 5\% of the entire population.
The data acquisition frequency of GPS locations varies according to the movement speed of the user to minimize the burden on the user's smartphone battery. 
If it is determined that the user is staying in a certain place for a long time, data is acquired at a relatively low frequency, and if it is determined that the user is moving, the data is acquired more frequently. 

\subsection*{Spatio-temporal Processing and Anonymization}
The set of mobile phone users included in the dataset were selected by spatially and temporally cropping the raw dataset. To spatially crop the raw dataset, we created a boundary box around an undisclosed metropolitan area in Japan, and selected mobile phone users who were observed within the boundary box more than 10 times during 10 day period (dates undisclosed for privacy reasons). 
To make the mobile phone users unidentifiable, the location pings are discretized into 500 meters x 500 meters grid cells and the timestamps into 30 minute bins. 
The actual date of the observations were also masked (i.e., timeslot $t$ of day $d$) to protect privacy.
The movement (encoded into 500m grid cells) of the mobile phone users was tracked across a total of 90 days (again, dates are undisclosed), including a 75-day period of business-as-usual (\textit{Dataset 1}) and another 15-day period under an emergency situation (\textit{Dataset 2}), where we can assume human behavior and mobility patterns could be shifted.  
The dataset was finally cropped by selecting users with a sufficient number of 30-minute timeslot observations to ensure that the mobility patterns could be studied (see Figure \ref{fig:user_pings_cells} for distribution of pings per user). Observations outside of the target boundary box were discarded. For \textit{Dataset 1}, 100,000 users were selected, and for \textit{Dataset 2}, 25,000 users were selected.

\subsection*{Privacy Policy}
Yahoo Japan Corporation (YJ) has developed its own privacy policy and requires users to read and agree to its privacy policy before using any of the services provided by YJ.
Furthermore, because location data is highly sensitive for the users, users were asked to sign an additional consent form specific to the collection and usage of location data when they used apps that collect location information. 
The additional consent explains the frequency and accuracy of location information collection, and also the purpose and how the data will be used.
Moreover, YJ implemented strict restrictions in the analysis procedure. 
The methodology for handling the data and for obtaining user consent for this study were supervised by an advisory board composed of external experts.
YJ also ensured that research institutions other than YJ that participate in this study (including co-investigators) do not have direct access to the data. 
Although external research institutions were allowed to analyze aggregated data, the actual raw data were kept within YJ, and any analysis performed on raw data were performed within servers administered by YJ. 
% In summary, the data collection and analysis were performed with careful consideration of the users' privacy.

\section*{Data Records}

\begin{figure}[t]
\centering
\includegraphics[width=.6\linewidth]{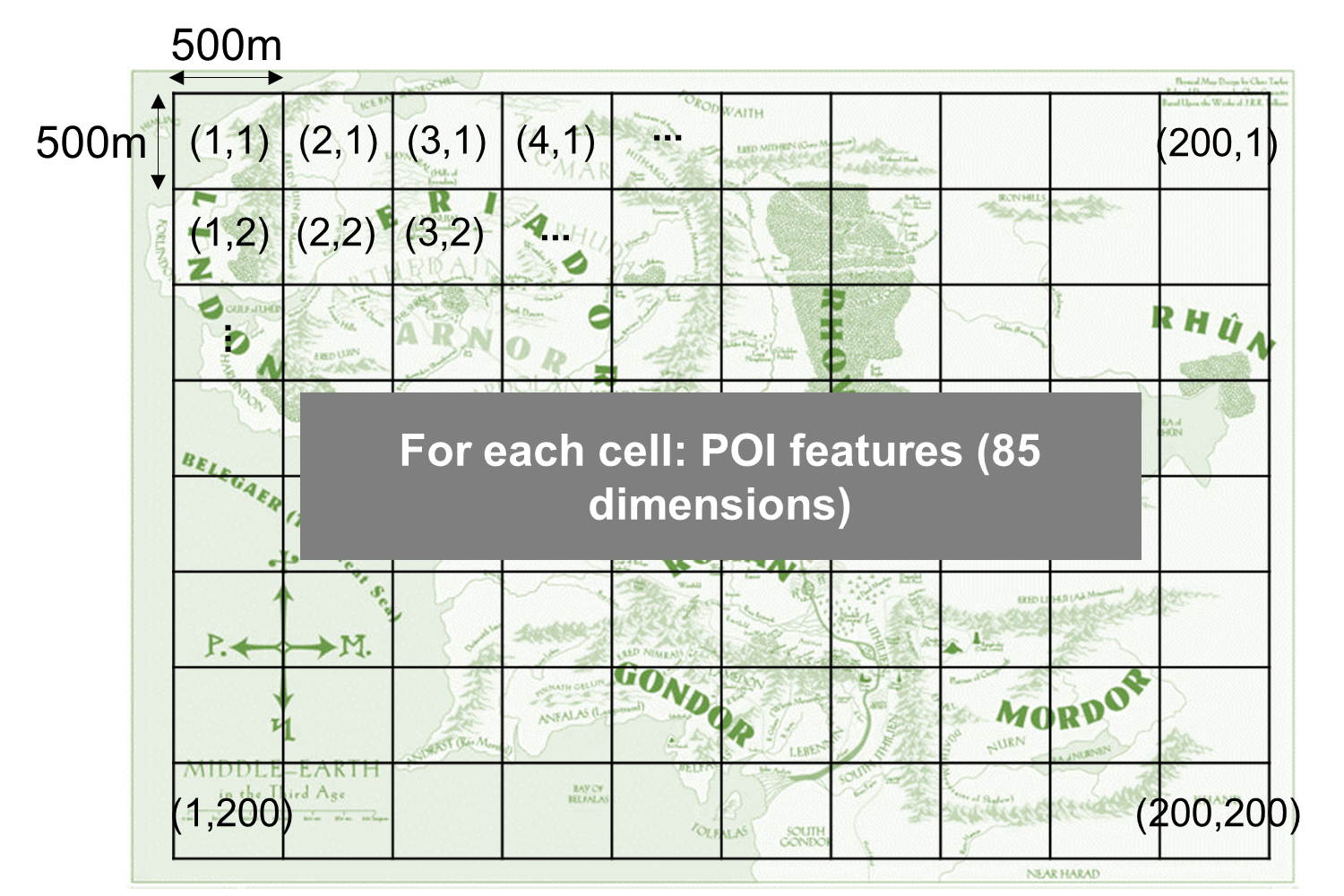}
\caption{Spatial layout of the target city and the grid cells. Each grid cell is approximately 500 meters x 500 meters, and the target area spans 200 x 200 grid cells.}
\label{fig:spatial}
\end{figure}

Table \ref{tab:example} shows an example of the dataset provided. Each record refers to an observation of an individual: 

\begin{itemize}
    \item \texttt{user ID} is the unique identifier of the mobile phone user (type: integer)
    \item \texttt{day} is the masked date of the observation. It may take a value between 0 and 74 for both Dataset 1 and Dataset 2 (type: integer). 
    \item \texttt{timeslot} is the timestamp of the observation discretized into 30 minute intervals. It may take a value between 0 and 47, where 0 indicates between 0AM and 0:30AM, and 13 would indicate the timeslot between 6:30AM and 7:00AM.
    \item \texttt{x,y} are the coordinates of the observed location mapped onto the 500 meter discretized grid cell. It may take a value between (1,1) and (200,200). Details are shown in Figure \ref{fig:spatial}. 
\end{itemize}

\begin{table}[t]
\centering
\begin{tabular}{ccccc}
\hline
user ID & day & timeslot & x & y \\
\hline
1 & 1 & 13 & 10 & 13 \\
1 & 1 & 18 & 11 & 15 \\
1 & 1 & 24 & 11 & 17 \\
1 & 1 & 27 & 12 & 19 \\
\multicolumn{5}{c}{...} \\
2 & 3 & 15 & 31 & 19 \\
2 & 3 & 28 & 35 & 33 \\
2 & 4 & 12 & 35 & 36 \\
\multicolumn{5}{c}{...} \\
\hline
\end{tabular}
\caption{\label{tab:example}Example of dataframe and the columns in the human mobility trajectory dataset.}
\end{table}

\section*{Basic Statistics of the Data}

\begin{figure}[h]
\centering
\includegraphics[width=.9\linewidth]{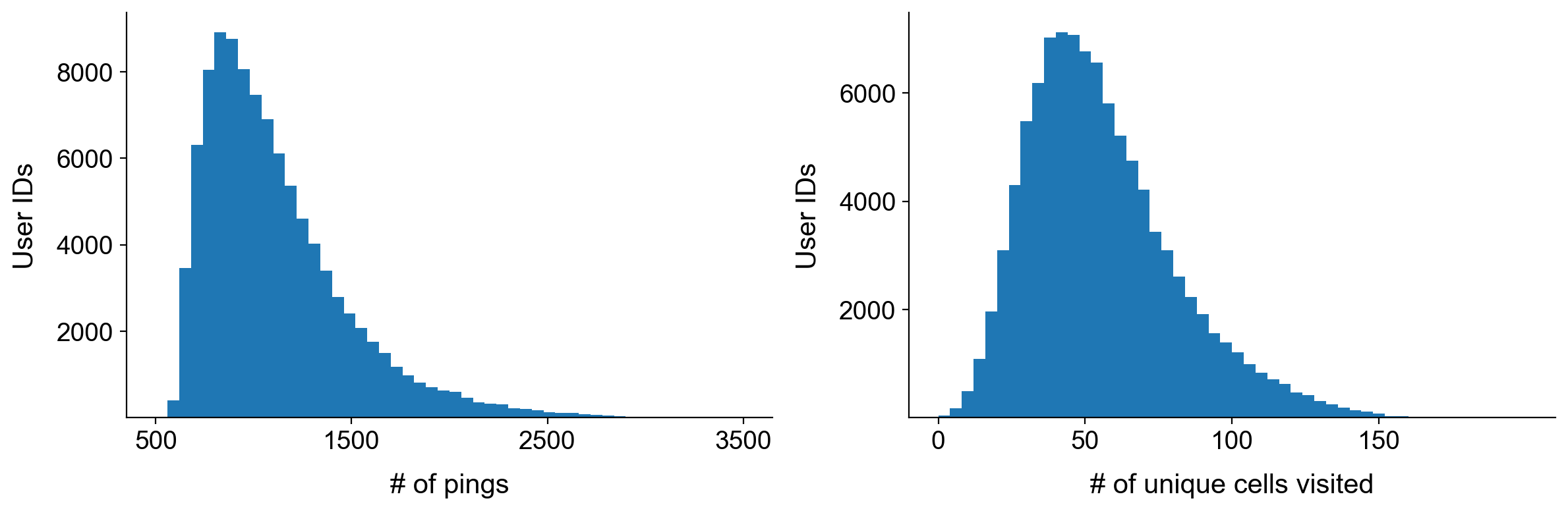}
\caption{Histograms of the number of GPS location data pings and number of unique cells visited per user, across the 75 day period stored in Dataset 1.}
\label{fig:user_pings_cells}
\end{figure}

\begin{figure}[h]
\centering
\includegraphics[width=.9\linewidth]{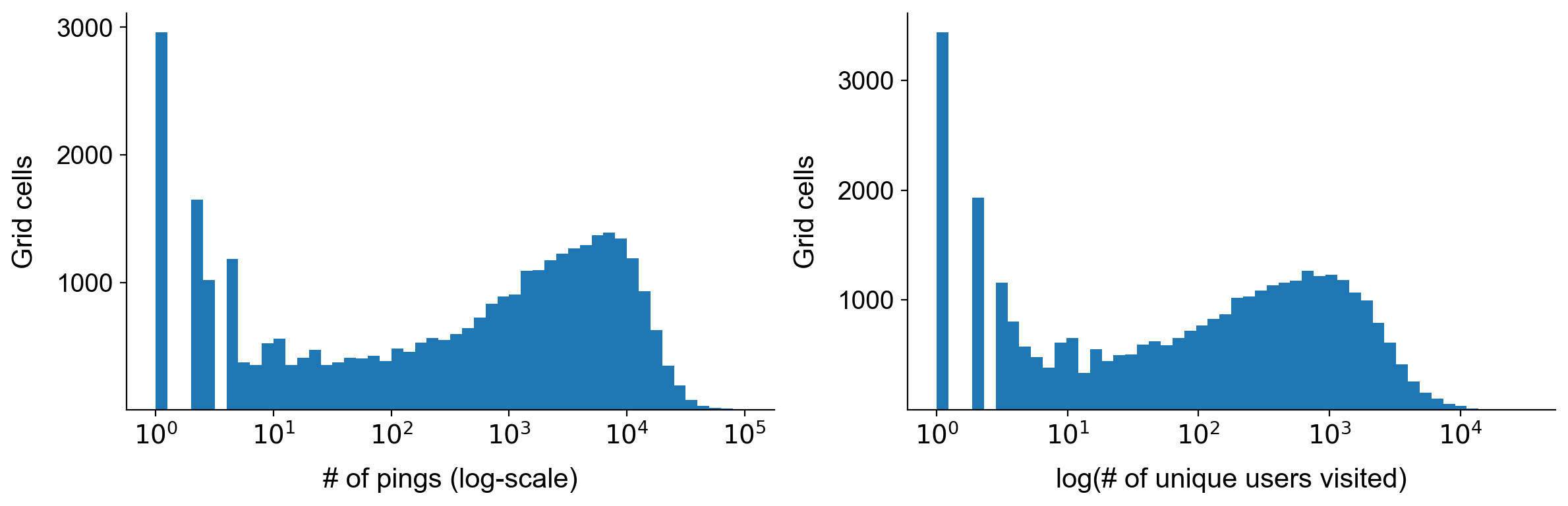}
\caption{Histograms of the number of GPS location data pings and number of unique users visited per grid cell, across the 75 day period stored in Dataset 1.}
\label{fig:cell_pings_uids}
\end{figure}

To provide guidance for data users, we have computed the basic descriptive statistics of \textit{Dataset 1}. 
The total number of records are 111,535,175, with exactly 100,000 unique users (numbered 0 to 99,999), across 75 days (numbered 0 to 74), in 48 different 30 minute timesteps (numbered 0 to 47). 
Figure \ref{fig:user_pings_cells} shows the histogram of the number of pings per user ID (left) and the number of unique cells visited per user ID. 
Both plots show a skewed distribution, where a small fraction of the users are observed many times (i.e., more than 2000 pings, at 100 unique cells). 
Figure \ref{fig:cell_pings_uids} shows the histogram of the number of pings per user ID (left) and the number of unique users visited to each grid cell. Note that the x-axis in both plots are log-scaled.
Both plots show a bimodal distribution, where a large fraction of the cells are visited very few times (less than 10 pings or unique users) while another mode can be observed at around 10000 pings and 1000 unique users visited. This highlights the mix of urban and rural areas in the target region.

\begin{figure}[h]
\centering
\includegraphics[width=.9\linewidth]{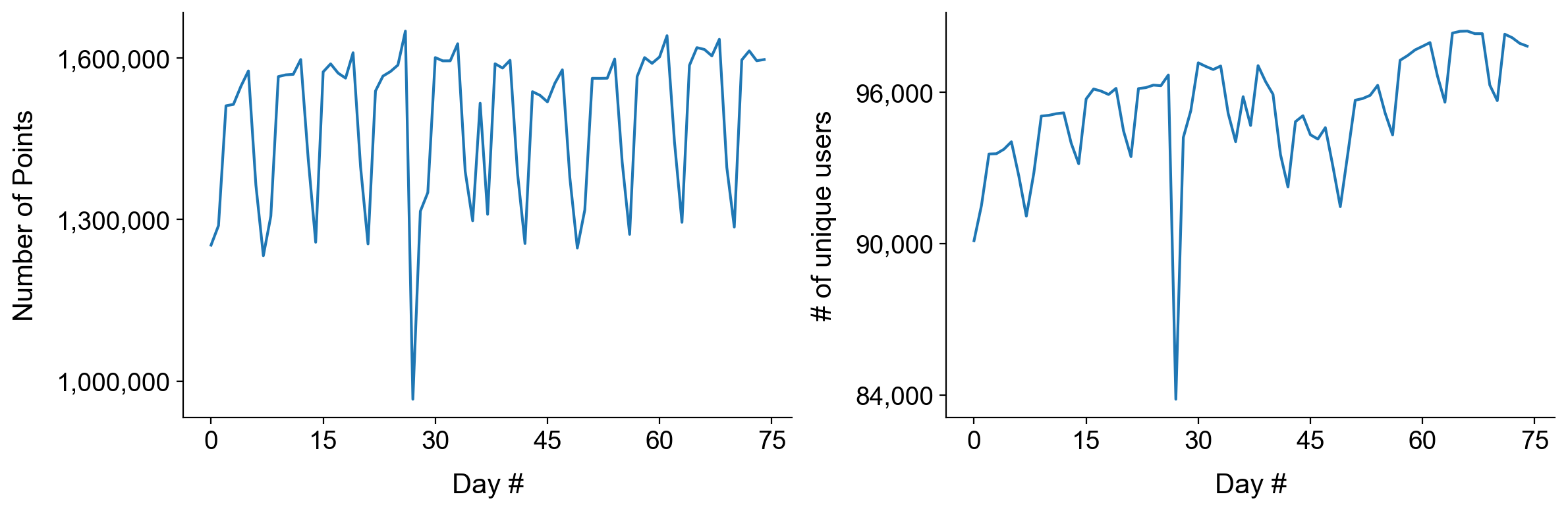}
\caption{Temporal dynamics of the number of pings and unique users per day (from day 0 to 74) in Dataset 1.}
\label{fig:days_pings_users}
\end{figure}

\begin{figure}[h]
\centering
\includegraphics[width=.9\linewidth]{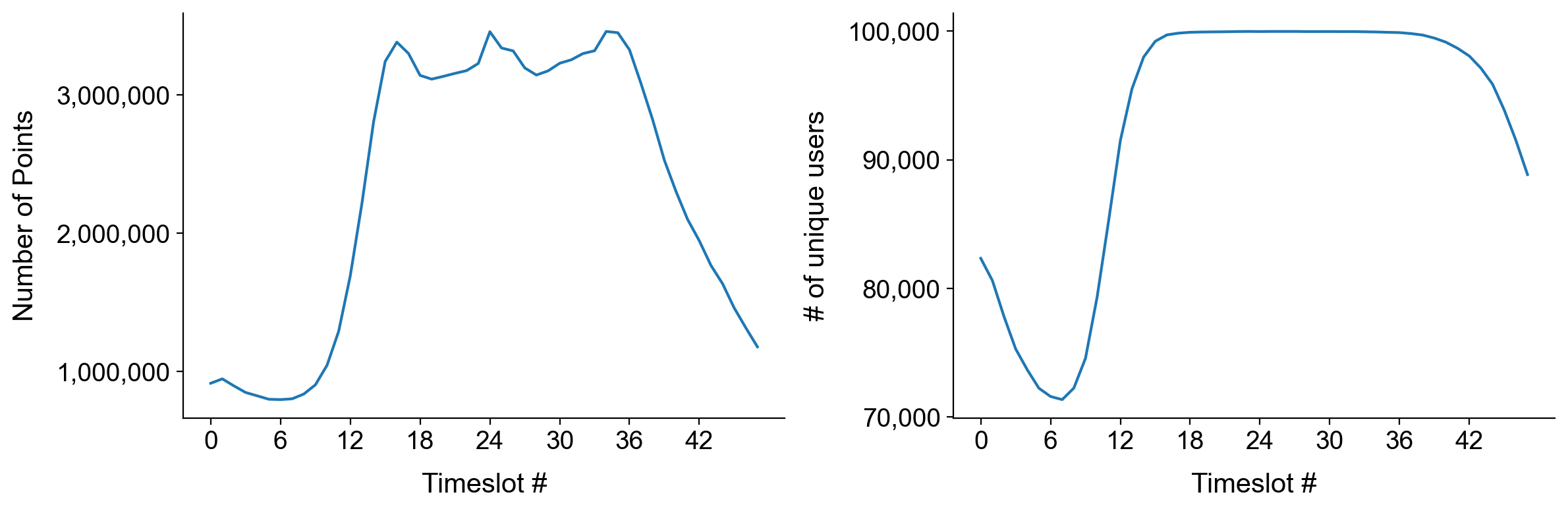}
\caption{Temporal dynamics of the number of pings and unique users per timeslot (from timeslot 0 to 47) in Dataset 1.}
\label{fig:time_pings_users}
\end{figure}

\begin{figure}[h]
\centering
\includegraphics[width=\linewidth]{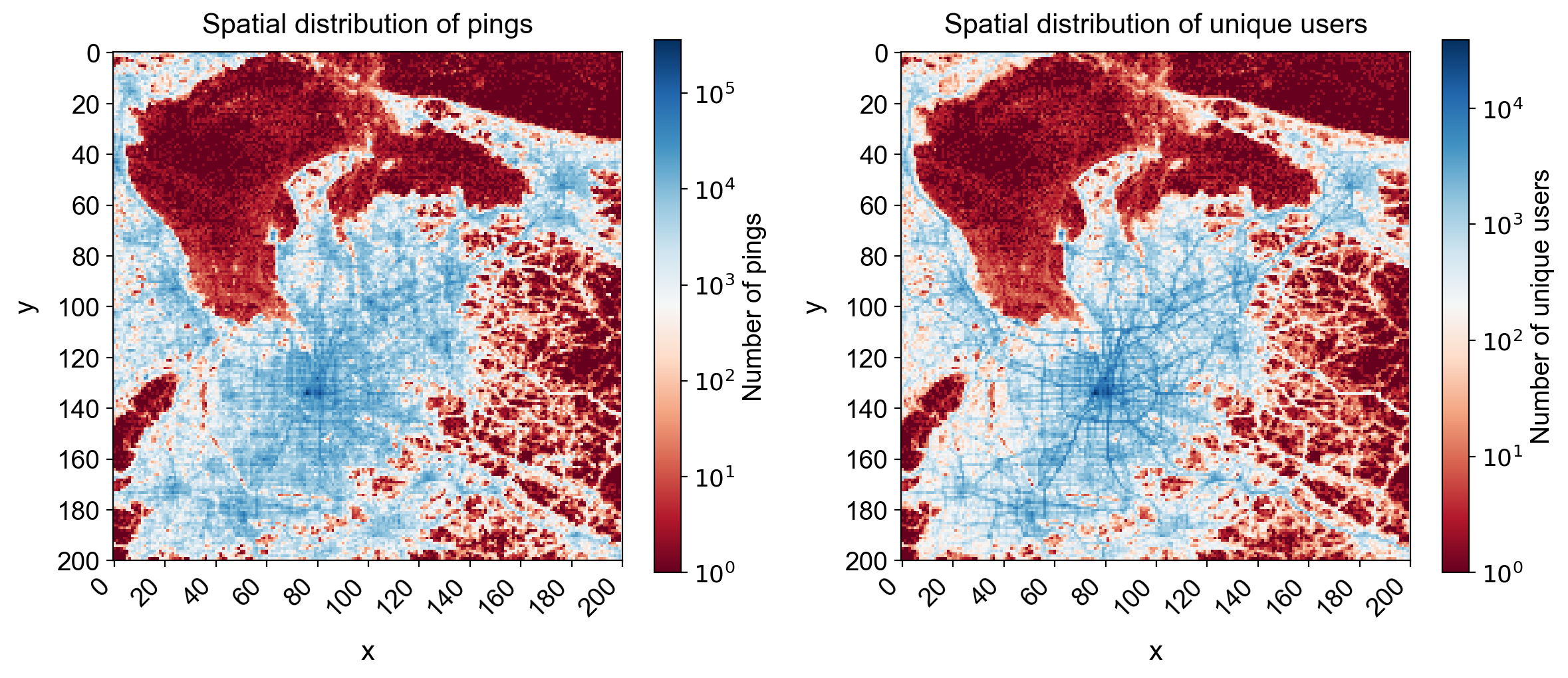}
\caption{2-dimensional histogram of the number of pings and the number of observed unique users across the 75 days. Note that the scales are log-scaled. The patterns show clear urban (blue) and rural (red) areas.}
\label{fig:map_pings_users}
\end{figure}

Figure \ref{fig:days_pings_users} shows the temporal dynamics of the number of pings and unique users per day (from day 0 to 74) in Dataset 1. The patterns show temporal regularity, showing clear patterns of weekdays and weekends. There is an anomaly on day 27, however this is due to a data collection issue. 
The unique number of users observed each day fluctuates more, showing a decrease near days 40 to 50 and an increase from day 60 onwards. 
Figure \ref{fig:time_pings_users} shows the temporal dynamics of the number of pings and unique users per timeslot (from timeslot 0 to 47) aggregated across all days observed in Dataset 1. The patterns show temporal regularity, showing clear morning and daytime peaks. 
The unique number of users observed between timeslot 12 (6AM) and timeslot 40 (8PM) is stable at around 100,000, showing a high observability during those time periods. 
Figure \ref{fig:map_pings_users} shows a 2-dimensional histogram of the number of pings and the number of observed unique users across the 75 days. Note that the scales are log-scaled. The patterns show clear urban (blue) and rural (red) areas.

% This section presents any experiments or analyses that are needed to support the technical quality of the dataset. This section may be supported by figures and tables, as needed. This is a required section; authors must present information justifying the reliability of their data.

\clearpage

\section*{Human Mobility Prediction Data Challenge}

The challenge takes place in a mid-sized and highly populated metropolitan area, somewhere in Japan. The area is divided into 500 meters x 500 meters cells, which span a 200 x 200 grid, as shown in Figure \ref{fig:spatial}. 
The human mobility datasets (`task1\_dataset.csv.gz' and `task2\_dataset.csv.gz') contain the movement of a total of 100,000 individuals across a 90 day period, discretized into 30-minute intervals and 500 meter grid cells. The first dataset contains the movement of a 75 day business-as-usual period, while the second dataset contains the movement of a 75 day period during an emergency with unusual behavior.

\begin{figure}[h]
\centering
\includegraphics[width=\linewidth]{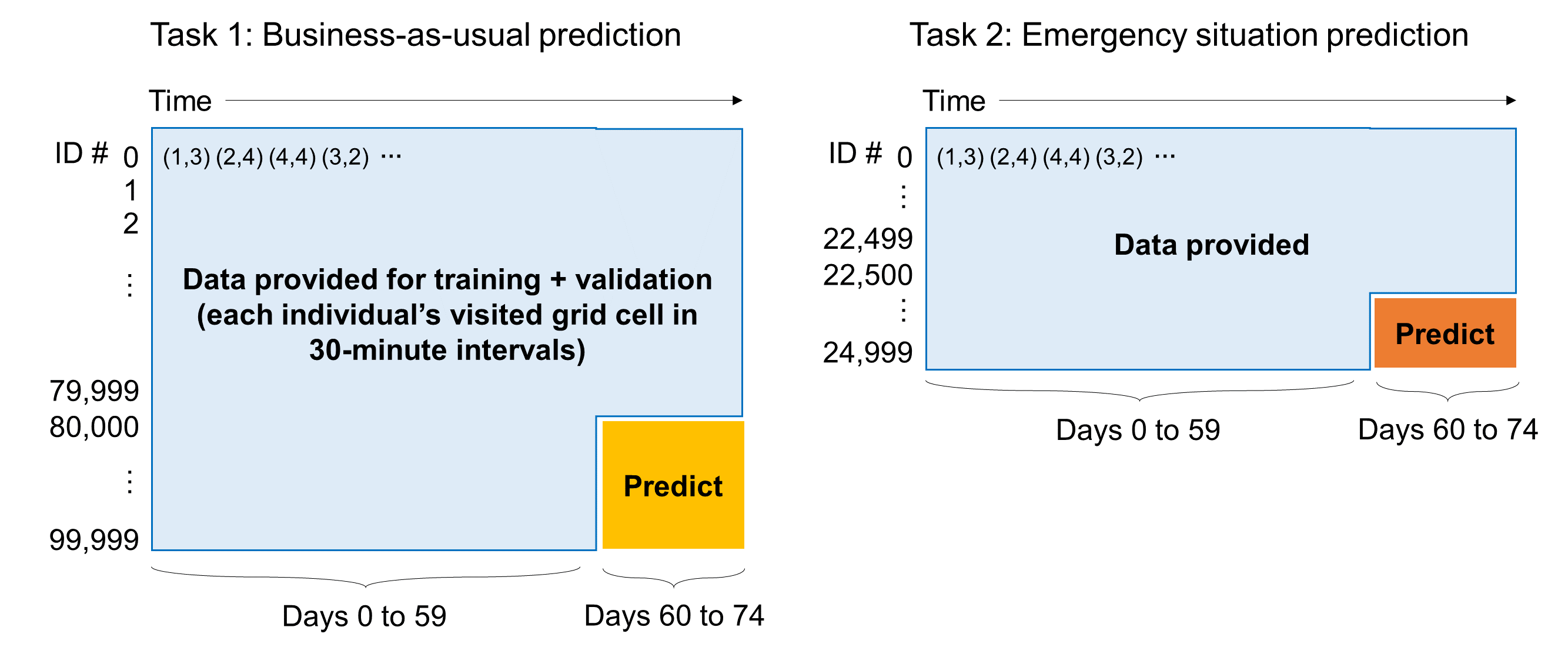}
\caption{2 tasks in the Human Mobility Prediction Challenge. 
In task 1, participants predict the movement of a subset (20,000) of the individuals for days 60 to 74 during a business-as-usual period. In task 2, participants predict the movement of a subset (2,500) of the individuals for days 60 to 74 during an emergency situation. }
\label{fig:tasks}
\end{figure}

There are 2 tasks in the Human Mobility Prediction Challenge, as shown in Figure \ref{fig:tasks}.
In task 1, participants are provided with the full time series data (75 days) for 80,000 individuals, and partial (only 60 days) time series movement data for the remaining 20,000 individuals (`task1\_dataset.csv.gz'). Given the provided data, Task 1 of the challenge is to predict the movement patterns of the individuals in the 20,000 individuals during days 60-74. Task 2 is similar task but uses a smaller dataset of 25,000 individuals in total, 2,500 of which have the locations during days 60-74 masked and need to be predicted (`task2\_dataset.csv.gz').

While the name or location of the city is not disclosed, the participants are provided with points-of-interest (POIs; e.g., restaurants, parks) data for each grid cell (~85 dimensional vector) as supplementary information (which is optional for use in the challenge) (`cell\_POIcat.csv.gz').
For more details, see \url{https://connection.mit.edu/humob-challenge-2023}.

\subsection*{Provided Datasets and Tasks}
The data challenge participants will be provided with 3 datasets -- HuMob datasets \#1 and \#2 (which are derived from the original human mobility dataset), and the POI dataset which may be used to supplement the prediction of human mobility. 

The data may be downloaded from \url{https://zenodo.org/record/8111993}. For teams to be granted access to the data, teams should request access via the Zenodo website by providing the name and email address of the lead investigator, and the following information in the `Justification' box: 
\begin{itemize}
    \item full list of members (name, institution, email address)
    \item team name (alphabets and numbers only, keep it $\leq 10$ characters)
\end{itemize}
Upon approval by the organizing team, the data will be available for download. 
If you do not receive the data approval within 24 business day hours, please contact \texttt{humob2023@gmail.com} with your information.

\textbf{Participants shall not carry out activities that involve unethical usage of the data, including attempts at re-identifying data subjects, harming individuals, or damaging companies. Participants will be allowed to submit full versions of their works to venues of their choice, upon approval by the organizers.
}

\subsubsection*{HuMob dataset \#1 (\texttt{task1\_dataset.csv.gz})}
Contains the movement of 100,000 individuals in total during a business-as-usual scenario. 80,000 individuals' movements are provided completely (for 75 days), and the remaining 20,000 individuals' movements for days 61 to 75 are masked as \texttt{999}. The challenge is to use the provided data to predict the masked cell coordinates (i.e., replace the `\texttt{999}'s). See Table \ref{tab:example} for the data format.

\subsubsection*{HuMob dataset \#2 (\texttt{task2\_dataset.csv.gz})}
Contains the movement of 25,000 individuals in total during a business-as-usual scenario. 22,500 individuals' movements are provided completely (for 75 days), and the remaining 2,500 individuals' movements for days 61 to 75 are masked as \texttt{999}. Similar to task 1, the challenge is to use the provided data to predict the masked movement coordinates (i.e., replace the `\texttt{999}'s). See Table \ref{tab:example} for the data format.

\subsubsection*{POI dataset (\texttt{cell\_POIcat.csv.gz})}
To aid the prediction task, we have prepared an auxiliary dataset that provides the count of different points-of-interest categories in each grid cell (e.g., restaurants, cafes, schools). However, to maintain anonymity of the location, we are not able to provide the actual category name that corresponds to each dimension. Therefore, each cell has a 85 dimensional vector, as shown in Table \ref{tab:poi}.

\begin{table}[h]
\centering
\begin{tabular}{cccc}
\hline
x & y & POI category (dim) & \# of POIs \\
\hline
1 & 1 & 13 & 10 \\
1 & 1 & 18 & 11  \\
1 & 1 & 24 & 11  \\
1 & 1 & 27 & 12  \\
\multicolumn{4}{c}{...} \\
2 & 2 & 15 & 31  \\
2 & 2 & 28 & 35  \\
2 & 2 & 12 & 35 \\
\multicolumn{4}{c}{...} \\
\hline
\end{tabular}
\caption{\label{tab:poi}Example of dataframe and the columns in the POI category dataset. First two columns show the x and y coordinates of the grid cell, third column denotes the dimension of the POI category (between 1 and 85), and the fourth column shows how many POIs of the POI category dimension located in the grid cell.}
\end{table}

% X=90.4km, Y=111km

\subsection*{Evaluation Metrics}
Two metrics will be used to measure the accuracy of the predicted mobility trajectories:
\begin{itemize}
    \item Dynamic Time Warping (DTW) \cite{senin2008dynamic}, for evaluating the similarity of trajectories as a whole, with step-by-step alignment.
    \item GEO-BLEU \cite{shimizu2022geo}, a metric with a stronger focus on local features, as in similarity measures for natural language sentences. Python implementation for the \texttt{GEOBLEU} metric can be found at \url{https://github.com/yahoojapan/geobleu}.
\end{itemize}

Submissions will be ranked for each metric, and the top 10 teams will be decided based on the two rankings. We recommend the teams to try to optimize for both metrics. 

\subsection*{Submission Procedure and Rules}

\begin{itemize}
    \item Prediction results for Tasks 1 and 2 should be uploaded to an online storage (e.g., Dropbox, Box, Google Drive, etc.) and the download links should be sent to \texttt{humob2023@gmail.com}. 
    \item The attached files should be named as: \texttt{{teamnumber}\_\{task1,task2\}\_humob.csv.gz}. For example, team number 5 submitting their solutions for task 1 should submit their prediction as \texttt{5\_task1\_humob.csv.gz}. 
    \item Only 1 submission per team would be evaluated. The final submission before the deadline (September 15th 23:59 AOE) will be considered as the final submission. 
    \item The format of the submission should include the same 5 columns as the original dataset (user ID, day, timeslot, x, y). Separate the columns using commas (,) and include no redundant spaces, and save the file using the \texttt{csv.gz} format.
    \item \emph{Only send the data for the predicted users}. For Task 1, only users \#80000 to \#99,999, and for Task 2, only users \#22500 to \#24999. 
\end{itemize}

\subsection*{Important Dates}
The top 10 teams with the best predictions will be invited to submit a final report with details of the methods and to present their work at the HuMob 2023 Workshop held in conjunction with ACM SIGSPATIAL 2023 in Hamburg, Germany on November 13th, 2023. We have prizes for the top 3 participants! 

\begin{itemize}
    \item June 15, 2023: data challenge announcement 
    \item July 10, 2023: data open at \url{https://zenodo.org/record/8111993}
    \item September 15, 2023: submission deadline for predictions
    \item September 22, 2023: notification of top contestants 
    \item October 14, 2023: submission deadline of workshop papers for top 10 teams 
    \item October 20, 2023: camera-ready submission 
    \item November 13, 2023: presentation in the workshop 
\end{itemize}

\subsection*{Organizing Team}
The team members are: Dr. Takahiro Yabe, MIT; Dr. Kota Tsubouchi, Yahoo Japan Corporation; Toru Shimizu, Yahoo Japan Corporation; Professor Yoshihide Sekimoto, University of Tokyo; Professor Kaoru Sezaki, University of Tokyo; Professor Esteban Moro, MIT; Professor Alex ‘Sandy’ Pentland, MIT. 
For general questions about the challenge: \texttt{humob2023@gmail.com}

% The Usage Notes should contain brief instructions to assist other researchers with reuse of the data. This may include discussion of software packages that are suitable for analysing the assay data files, suggested downstream processing steps (e.g. normalization, etc.), or tips for integrating or comparing the data records with other datasets. Authors are encouraged to provide code, programs or data-processing workflows if they may help others understand or use the data. Please see our code availability policy for advice on supplying custom code alongside Data Descriptor manuscripts.

% For studies involving privacy or safety controls on public access to the data, this section should describe in detail these controls, including how authors can apply to access the data, what criteria will be used to determine who may access the data, and any limitations on data use. 

\section*{Code availability}
The dataset can be downloaded from \url{https://zenodo.org/record/8111993}, and details about the Data Challenge can be found in \url{https://connection.mit.edu/humob-challenge-2023}.
Python implementation for the \texttt{GEOBLEU} metric can be found at \url{https://github.com/yahoojapan/geobleu}.

\bibliography{sample}

% \noindent LaTeX formats citations and references automatically using the bibliography records in your .bib file, which you can edit via the project menu. Use the cite command for an inline citation, e.g. \cite{Kaufman2020, Figueredo:2009dg, Babichev2002, behringer2014manipulating}. For data citations of datasets uploaded to e.g. \emph{figshare}, please use the \verb|howpublished| option in the bib entry to specify the platform and the link, as in the \verb|Hao:gidmaps:2014| example in the sample bibliography file. For journal articles, DOIs should be included for works in press that do not yet have volume or page numbers. For other journal articles, DOIs should be included uniformly for all articles or not at all. We recommend that you encode all DOIs in your bibtex database as full URLs, e.g. https://doi.org/10.1007/s12110-009-9068-2.

% \section*{Acknowledgements}
% Acknowledgements should be brief, and should not include thanks to anonymous referees and editors, or effusive comments. Grant or contribution numbers may be acknowledged.

\section*{Author contributions statement}

T.Y. and K.T developed and computed the mobility indices. T.S. developed the evaluation metric code. All authors wrote and reviewed the manuscript. 

\section*{Competing interests} 
% (mandatory statement)
% The corresponding author is responsible for providing a \href{https://www.nature.com/sdata/policies/editorial-and-publishing-policies#competing}{competing interests statement} on behalf of all authors of the paper. This statement must be included in the submitted article file.
The authors declare no competing interests.

\end{document}